\documentstyle[12pt]{article}
\input epsf
\begin{document}
\def \beq{\begin{equation}}
\def \eeq{\end{equation}}
\rightline{EFI-97-10}
\rightline{hep-ex/9702008}
\bigskip
\centerline{\bf DETECTION OF THE RF PULSE}
\centerline{\bf ASSOCIATED WITH COSMIC RAY AIR SHOWERS
\footnote{Based on an appendix to proposal for the Auger Air Shower Array}}
\bigskip
\centerline{\it Jonathan L. Rosner}
\centerline{\it Enrico Fermi Institute and Department of Physics}
\centerline{\it University of Chicago, Chicago, IL 60637}
\centerline{and}
\centerline{\it John F. Wilkerson}
\centerline{\it Department of Physics}
\centerline{\it University of Washington, Seattle, WA 98195}
\bigskip

\centerline{\bf ABSTRACT}
\medskip
\begin{quote}
Initial results of a project to detect the radio-frequency pulse associated
with extensive air showers of cosmic rays are described briefly.  This work is
being performed at the CASA/MIA array in Utah, with the intention of designing
equipment that can be used in conjunction with the Auger Giant Array proposal. 
\end{quote}

\section{Motivation}

As a result of work in the 1960's and 1970's \cite{Chac,Allan}, some of which
has continued beyond then (see, e.g., \cite{Agasa,Eastop,Yakutsk,VHF}), it is
recognized that air showers of energy 10$^{17}$ eV are accompanied by
radio-frequency pulses, whose polarization and frequency spectrum suggest that
they are due mainly to the separation of positive and negative charges of the
shower in the Earth's magnetic field \cite{KL}.  The most convincing data have
been accumulated in the 30--100 MHz frequency range. However, opinions have
differed regarding the strength of the pulses, and atmospheric and ionospheric
effects have led to irreproducibility of results. In particular, there may also
be pulses associated with cosmic-ray-induced atmospheric discharges \cite{atm}.
There are reports of detection at MHz or sub-MHz frequencies
\cite{Agasa,Eastop,Yakutsk}, which could be associated with such a mechanism.
Signals above 100 MHz have also been reported \cite{VHF}. 

A study is being undertaken of the feasibility of equipping the Auger array
with the ability to detect such pulses.  It is possible that the higher energy
of the showers to which the array would be sensitive would change the
parameters of detection.  Before a design for large-scale RF pulse detection
can be produced, it has been necessary to retrace some of the steps of the past
30 years by demonstrating the existence of the pulses for 10$^{17}$ eV showers,
and by controlling or monitoring some of the factors which led to their
irreproducibility in the past.  RF pulses may be able to provide auxiliary
information about primary composition and shower height \cite{Allan}.

In this report we describe the prototype activity at the CASA/MIA site, note
related activities, and set forth some considerations regarding plans for the
Auger project.  More concrete plans for RF detection at Auger must await the
outcome of the ongoing work at the CASA/MIA site. 

\section{CASA/MIA Prototype setup}

In order to verify the claim \cite{Chac,Allan} that 10$^{17}$ eV showers are
accompanied by RF pulses with significant energy in the 30--100 MHz range, a
prototype detector has been set up at the CASA/MIA site in Dugway, Utah.  This
section describes the status of that effort. 

\subsection{Large-event trigger}

A trigger based on the coincidence of several muon ``patches'' was set to
select ``large'' showers with a rate of 30 -- 50 per hour \cite{mutrig}. The
MIA Patch-Sum trigger was sent into a fan-in/fan-out.  From this, the signal
was put into a LeCroy 821 Discriminator, amplified, and sent over a cable to a
trailer located just to the east of the CASA array.  The location of the
trailer outside the array was dictated by the intense RF noise within the array
determined by surveys taken in 1995. 

The electronics was set to trigger at 7 of the eight outermost muon patches.
This was estimated to correspond to a minimum shower energy of about $2 \times
10^{16}$ eV, based on the rate at $10^{18}$ eV of 1/km$^2$/day/sr.  At this
level good correlation could be established between trigger pulses and events
recorded by the CASA data acquisition system. 

\subsection{Monitoring of RF noise environment}

It was a concern that the RF noise of the local electronics and the presence of
an extensive lightning-protection array might dictate the placement of the RF
pulse detection system outside the periphery of the array. The behavior of a
single CASA board was investigated at the University of Chicago. The various
clock signals were detected at short distances ($< 1$ m) from the board, but a
much more intense set of harmonics of 78 kHz emanated from the switching power
supplies.  These harmonics persisted well above 100 MHz.  At 144--148 MHz, they
overlapped, leading to intense broad-band noise.

A spectrum analyzer, obtained from the now-defunct SSC Laboratory, was used to
make a broad survey of the RF noise at the CASA site in various frequency
ranges and at various locations.  Surveys indicated considerable noise within
the array, but a much lower level sitting just outside the array. Consequently,
it was decided to set up a special trailer for monitoring RF pulses just to
the east of the array.

\subsection{Data acquisition}

A log-periodic antenna sensitive to 26 -- 170 MHz was mounted about 30 meters
to the east of the CASA boundary, at an elevation of about 10 meters to place
it just above the lightning protection grid.  A digital storage scope was used
to register filtered and preamplified RF data on a rolling basis. These data
were then captured and stored upon receipt of a large-event trigger. 

The experiment was repeated using successively greater amounts of amplification
and narrower band-pass filters.  Data were taken using two systems on two
separate computers (at different times), allowing for analysis both by
Wilkerson and students at the University of Washington and by Rosner at
Chicago.  Wilkerson's data are still being processed, so the preliminary
results to be mentioned below will be based on Rosner's analysis. 

\subsection{Results}

The most recent configuration involved feeding the signal from the antenna
through Mini-Circuits BHP-25 and BLP-30 filters, with 6 dB points of 23 and 39
MHz, a Mini-Circuits ZFL-500LN preamplifier  with 26 dB of gain, a
Mini-Circuits BBP-30 bandpass filter with 6 dB points 24.5 and 37 MHz, and
another ZFL-500LN preamplifier. This configuration was arrived upon after
numerous runs in which efforts were made to minimize noise from sources such as
computers, monitors, and the preamplifiers themselves.  It was employed for
an 8-1/2 hour run during which 305 triggers were received.  Each trigger caused
50 $\mu$s of RF data, centered around the trigger and acquired at 1 GSa/s,
to be saved on a Tektronix TDS540B digitizing oscilloscope.  These data were
then passed using a GPIB interface to a Dell XPS200s Pentium computer, where
they were stored for future analysis.

Events were scanned for large positive peaks. Three points each exceeding 15.2
mV on the oscilloscope input were demanded within a window of 10 ns, in order
to eliminate both constant background noise and very-high-frequency transients.
If an event contained such a peak, the largest peak within an associated group
of duration less than 300 ns was then identified and its height was plotted as
a function of time relative to the trigger pulse. The results are shown in
Fig.~1(a). 

A clear signal is visible in the window between $-10$ and $-5~\mu$s relative
to the trigger.  The trigger pulse itself requires several $\mu$s to be formed
at the CASA site, and an additional 2.15 $\mu$s to propagate to the RF
detection trailer.  Consequently, the signal is likely to be due to the prompt
RF signal associated with showers. 

\begin{figure}
\centerline{\epsfysize = 2.5in \epsffile {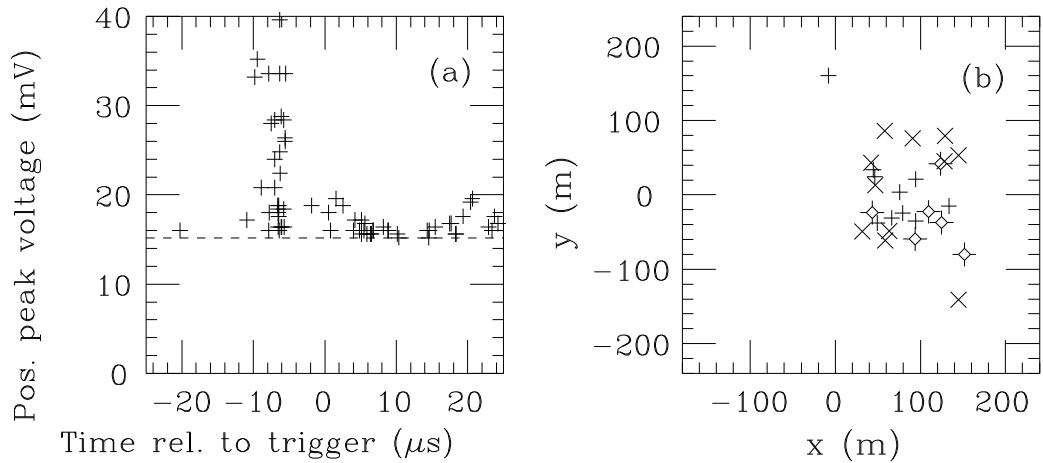}}
\caption{(a) Maximum pulse height as a function of arrival time relative to
trigger. (b) Distribution of cores with respect to center of CASA array for
events in (a) between $10$ and $-5~\mu$s. $\times:~ 15.2 < V_{\rm pk} < 20$ mV;
$+:~20 \le V_{\rm pk} < 30$ mV; fancy points: $30$ mV $\le V_{\rm pk}$. Here
$(x,y) = $ core position (east, north) of array center.  The antenna was
located at $(x,y) \simeq (270,0)$ m.} 
\end{figure}

The pulses within the window from $-10$ to $-5~\mu$s in Fig.~1(a) are
associated with showers whose cores nearly all lie east of the center of the
CASA array, as shown in Fig.~1(b).  This result is in accord with the rapid
decrease of pulse intensity with distance from the core observed by the Haverah
Park group \cite{Allan}. 

\subsection{Near-term plans}

A total of more than 5000 triggers, obtained under various conditions of
filtering, preamplification, and noise reduction, remains to be analyzed.
Preliminary results from runs in September and December of 1996 indicate
that about 1 in 10 of these triggers may contain information on RF pulses
from air showers.  Shower sizes, core locations, and incident shower angles
will be correlated with time, duration, and intensity of radio bursts.  Data
must also be collected with the CASA detector disabled (leaving only the
MIA trigger), in order to exclude the possibility that signals are due to
pulses involved in CASA data transfer.

A non-imagining Cherenkov light array (BLANCA) has recently been added at the
CASA site.  RF data will be collected during at least one period when BLANCA
is active in order to utilize the additional information that BLANCA may
provide on shower energy and primary composition.

Still to be performed are experiments which seek to monitor RF pulses at lower
frequencies and at greater distances from the array. For these pulses, whose
strengths may be correlated with atmospheric electric fields, it is planned to
monitor such fields with the help of a field mill, which Rosner has recently
constructed. A group working at the AGASA array in Japan \cite{Agasa} has
detected pulses associated with large air showers (typically above $10^{18}$
eV) at frequencies below 500 kHz.  These pulses are of several microseconds
duration.  While they can be associated with cores more than 1 km away from the
antenna, they do not always occur, even for very large showers.  It is
suspected that they may be associated with an atmospheric discharge mechanism. 
Consequently, once suitable low-frequency filters are found to eliminate
signals from the AM broadcast band and the 284 kHz air traffic control beacon
at Dugway, a survey of pulses below 500 kHz will be undertaken at Dugway. 
Simultaneous monitoring of electrostatic atmospheric electric field gradients
at ground level will provide at least a first look at the possibility that a
discharge mechanism is involved. 

Wilkerson was engaged in projects at Los Alamos whose aim was detection of
electromagnetic pulses, including those possibly produced by cosmic-ray-induced
electromagnetic discharges, with frequencies in the 30 -- 100 MHz range.  Part
of this work included building hardware for self-triggering on short duration
wide-band RF pulses.  Many of the pulse identification, fast-digitization and
memory problems are identical to those for pulse detection at CASA/MIA. 
Time-frequency plots have been obtained which are exactly those one would
generate in a survey at CASA/MIA. Wilkerson has also encountered similar
requirements for digitization of SNO data.  His estimate is that one can use
Maxim MAX 100 A/D chips for less than \$1K per channel, but that feeding their
output into memory may well amount to another \$1K per channel. Other
references on digitizers have been obtained \cite{Atiya,Bryman}. 

Wilkerson has moved his data acquisition system to CASA and has taken data from
the digitizing scope for analysis at the University of Washington.
Time-frequency plots obtained for a day's data in September, 1996, already
showed behavior consistent with Fig.~1.  In the near future the RF pulse
detection system will be activated with a suitable set of filters, in order to
detect pulses from air showers without the need for an external trigger. 

In the future the spectrum analyzer obtained by Rosner may of help in detecting
potential sources of interference to RF communications in the Auger project. 

\section{Related activities}

\subsection{Status of GHz detection}

David Wilkinson, who visited the University of Chicago during the spring of
1995, has promised to look into the power radiated at frequencies of several
GHz, where new opportunities exist associated with the availability of
low-noise receivers.  These techniques have now been implemented in the
RICE project \cite{RICE}, which seeks to detect pulses with 
frequency components around 250 MHz in Antarctic polar ice.

\subsection{Other options}

Dispersion between arrival times of GPS signals on two different frequencies
may serve as a useful monitor of air shower activity.  The possibility of
correlation of large showers with such dispersion events will be investigated. 

It may be possible at the CASA/MIA site to monitor commercial broadcast
signals in the 55 - 88 MHz range to detect momentary enhancements associated
with large showers, in the same sense that meteor showers produce such
enhancements.  Television Channels 3 and 6, for which no nearby stations
exist, offer one possibility.

\section{Considerations for Auger project}

At present we can only present a rough sketch of criteria for detection in the
30--100 MHz range. Data would be digitized at a 500 MHz rate at each station
and stored in a rolling manner, with at least 20 microseconds of data in the
pipeline at any moment.  Upon receipt of a trigger signaling the presence of a
``large'' shower ($> 10^{18}$ eV), these data would be merged into the rest of
the data stream at each station. 

Per station, we estimate the following additional costs, in US dollars,
for RF pulse detection:

\begin{center}
\begin{tabular}{l l c} \hline
Two antennas and protection circuitry:   & 200 & (a) \\
Mounting hardware:                       & 100 & (b) \\
Cables and connectors:                   & 200 & (c) \\
Preamps:                                 & 500 & (d) \\
Digitization and memory electronics:     &2000 & (e) \\ \hline
Total per station:                       &3000 & (f) \\ \hline
\end{tabular}
\end{center}

\noindent

\noindent
(a) Two military-surplus log-periodic antennas;
crossed polarizations.  Difference signal to be detected.

\noindent
(b) Highly dependent on other installations at site.  Antennas are
to be pointed vertically but optimum elevation not yet determined.

\noindent
(c) Antennas are mounted near central data acquisition site of each station,
but sufficiently far from any sources of RF interference such as switching
power supplies.

\noindent
(d) Commercial GaAsFET preamps and gas discharge tubes.

\noindent
(e) Subject to prototype development experience.  Power requirements
not yet known.

\noindent
(f) The number of stations to be equipped with RF detection will
depend on prototype experience.
\bigskip

The above estimate assumes that one can power the preamps and DAQ electronics
from the supply at each station without substantial added cost.  It also
assumes that a ``large-event trigger'' will be available at each station.
A further assumption is that the difference signal suffices to characterize
the pulse.  Additional preamplification and DAQ electronics may be required if
this is not so. One consideration may be the acquisition of antennas robust
enough to withstand extreme weather (particularly wind) conditions. 

For detection at frequencies above or below 30--100 MHz, the criteria are not
yet well enough developed to permit any cost estimate.

\section{Acknowledgments}

We thank Kevin Green for setting up and studying the ``large-event trigger,''
Rachel Gall for assisting him in this work during the summer of 1996, and Mike
Cassidy for help in all stages of the experiment. Funds provided by the Enrico
Fermi Institute and by the Physics Department of the University of Chicago were
instrumental in allowing this work the be undertaken. Thanks are also due to
Jim Cronin, Lucy Fortson, Dick Gustafson, Gerard Jendraszkiewicz, Larry Jones,
Brian Newport, Rene Ong, Dave Peterson, Leslie Rosenberg, Dave Smith, Augustine
M. Urbas, and Dave Wilkinson for help in various phases of this work.  J. L. R.
is grateful to M. Nagano for hospitality at the Institute for Cosmic Ray
Research in Tokyo, and to him and K. Kadota, F. Kakimoto, and M. Teshima for
discussions. This work was supported in part by the Louis Block Fund of the
University of Chicago.

\end{document}